\documentstyle[12pt]{article}

\def\thebibliography#1{\section*{
References}\list
  {\arabic{enumi}.}{\settowidth\labelwidth{#1}\leftmargin\labelwidth
    \advance\leftmargin\labelsep
    \usecounter{enumi}}
    \def\newblock{\hskip .11em plus .33em minus .07em}
    \sloppy\clubpenalty4000\widowpenalty4000
    \sfcode`\.=1000\relax}

\def\op#1{\mathop{\fam0 #1}\limits}

\newcommand{\pr}{{\rm pr}}
\newcommand{\beq}{\begin{equation}}
\newcommand{\eeq}{\end{equation}}
\newcommand{\ben}{\begin{eqnarray}}
\newcommand{\een}{\end{eqnarray}}
\newcommand{\be}{\begin{eqnarray*}}
\newcommand{\ee}{\end{eqnarray*}}
\newcommand{\bea}{\begin{eqalph}}
\newcommand{\eea}{\end{eqalph}}

\newcommand{\nm}[1]{\mid {#1}\mid}

\newcommand{\cP}{{\cal P}}
\newcommand{\cB}{{\cal B}}
\newcommand{\bL}{{\bf L}}
\newcommand{\bR}{{\bf R}}
\newcommand{\bZ}{{\bf Z}}

\newcommand{\La}{\Lambda}
\newcommand{\f}{\phi}

\newcommand{\G}{\Gamma}

\newcommand{\si}{\sigma}
\newcommand{\w}{\wedge}

\newcommand{\ol}{\overline}
\newcommand{\dr}{\partial}
\newcommand{\ot}{\otimes}

\newenvironment{eqalph}{\stepcounter{equation}
\setcounter{equationa}{\value{equation}}
\setcounter{equation}{0}

\begin{eqnarray}}{\end{eqnarray}
\setcounter{equation}{\value{equationa}}}

\newcommand{\nw}[1]{[{#1}]}
\newcommand{\der}{\rm Der}

\begin{document}
\hbox{}

\begin{center}
{\large \bf ON THE  GEOMETRIC ARENA OF SUPERMECHANICS}
\bigskip

{\sc G.Sardanashvily}
\bigskip

Department of Theoretical Physics, Moscow State University

117234 Moscow, Russia

E-mail: sard@grav.phys.msu.su 
\end{center}

\begin{abstract}
In the case of simple graded manifolds utilized in supermechanics,
supervector fields and exterior superforms are represented by global
sections of smooth vector bundles.
\end{abstract}
\bigskip

A BRST extension of Hamiltonian mechanics \cite{gozz,book98} shows that  (i)
one should consider vector bundles in order to introduce generators of BRST
and anti-BRST transformations, and (ii) one can narrow the
class of superfunctions under consideration because the BRST extension
of a Hamiltonian is a polynomial in odd variables.

The main ingredient in a theory of supermanifolds is the sheaf $\cB$ of
graded commutative algebras on a manifold $Z$. If this sheaf fulfills
the  Rothstein axioms, it is said to be an $R$-supermanifold
\cite{bart,cia}. In particular, the notion of an $R$-supermanifold includes
graded manifolds, supermanifolds of A.Rogers and
infinite-dimensional supermanifolds of A.Jadczyk and K.Pilch.
This notion also
implies that the sheaf $\der\, \cB$ of graded differentiations of $\cB$ and 
the dual sheaf $\der^*\cB$ are introduced.
Let $U$ be an open subset of $Z$ and $\cB\mid_U$ the restriction of the sheaf
 $\cB$ to $U$. By a graded differentiation of the sheaf
$\cB\mid_U$ is meant its endomorphism $u$ such that
\be
 u(ff')=u(f)f'+(-1)^{\nw u\nw f}fu (f') 
\ee
for the homogeneous elements $u\in\der\, \cB$ and $f,f'\in \cB\mid_U$.
We will use the
notation $\nw .$ of the Grassman parity. 
The graded differentiations of $\cB\mid_U$ constitute the $\cB$-module
$\der\,\cB(U)$, and the presheaf of these $\cB$-modules generates the sheaf
$\der\,\cB$. Its elements are 
called  supervector fields on a manifold $Z$.
The dual of the sheaf $\der\,\cB$ is the sheaf
 $\der^*\cB$ generated by the $\cB$-linear morphisms
\beq
\f:\der\cB(U)\to \cB_U. \label{z789}
\eeq
One can think of its elements as being 1-superforms on a manifold $Z$. 

We will consider the following class of graded manifolds, called simple graded
manifolds \cite{cari97}. 
Given a vector bundle $E\to Z$ with an $m$-dimensional
typical fiber $V$, let us consider the fiber bundle
\beq
\w E^*=\bR\op\oplus_Z(\op\oplus_{k=1}^m\op\w^k E^*) \label{z780}
\eeq
whose typical fiber is the finite Grassman algebra $\La^*=\w V^*$. 
Note that
there is a different notion of a Grassman algebra \cite{jad}, which is not 
equivalent to the above one if $V$ is finite-dimensional \cite{cia}.
Global sections of $\w E^*\to Z$ (\ref{z780}), called superfunctions, make up a
$\bZ_2$-graded ring $\cB^0$.  
Let $\{c^a\}$ be the holonomic bases in $E^*\to Z$ with respect to some bundle
atlas with transition functions $\{\rho^a_b\}$, i.e., $c'^a=\rho^a_b(z)c^b$.
Then superfunctions read
\beq
f=\op\sum_{k=0}^m \frac1{k!}f_{a_1\ldots
a_k}c^{a_1}\cdots c^{a_k}, \label{z785}
\eeq
where $f_{a_1\cdots
a_k}$ are local functions on $Z$, and we omit the symbol of exterior product of
elements $c$. The coordinate transformation law of superfunctions
(\ref{z785}) is obvious. The sheaf of these superfunctions 
belongs to the above mentioned class of simple graded manifolds. 
In this case, supervector fields and exterior superforms on $Z$ can be
represented by sections of smooth vector bundles over $Z$ as follows. 

Since the canonical splitting $VE= E\times E$, the vertical tangent bundle 
$VE\to E$ can be provided with the fiber bases $\{\dr_a\}$ dual of $\{c^a\}$.
These are fiber bases of $\pr_2VE=E$. Let $(z^A)$ be coordinates on $Z$. Then
supervector fields on a manifold $Z$ read
\beq
u= u^A\dr_A + u^a\dr_a  \label{z791}
\eeq
where $u^A, u^a$ are local superfunctions (\ref{z785}). A supervector field
$u$ (\ref{z791}) is homogeneous iff
\be
\nw{u^A}=\nw{u^B}=\nw{u^a}+1=\nw{u^b}+1=\nw u, \qquad \forall A,B,a,b.
\ee
It defines the graded endomorphism of $\cB^0$ by the law
\beq
u(f_{a\ldots}c^a\cdots)=u^A\dr_A(f_{a\ldots})c^a\cdots +u^a
f_{a\ldots}\dr_a\rfloor (c^a\cdots). \label{cmp50}
\eeq
A direct computation shows that this law implies the corresponding
coordinate transformation law 
\be
u'^A =u^A, \qquad u^a=\rho^a_ju^j +u^A\dr_A(\rho^a_j)c^j
\ee
of supervector fields. It follows that supervector fields can be represented
by sections of the vector bundle $\cP_E\to Z$ which is locally isomorphic to
the vector bundle
\be
\cP_E\mid_U\approx\w E^*\op\ot_Z(\pr_2VE\op\oplus_Z TZ)\mid_U,
\ee
and has the transition functions
\be
&& z'^A_{i_1\ldots i_k}=\rho^{-1}{}_{i_1}^{a_1}\cdots
\rho^{-1}{}_{i_k}^{a_k} z^A_{a_1\ldots a_k}, \\
&& v'^i_{j_1\ldots j_k}=\rho^{-1}{}_{j_1}^{b_1}\cdots
\rho^{-1}{}_{j_k}^{b_k}\left[\rho^i_jv^j_{b_1\ldots b_k}+ \frac{k!}{(k-1)!} 
z^A_{b_1\ldots b_{k-1}}\dr_A(\rho^i_{b_k})\right] 
\ee
of the bundle coordinates $(z^A_{a_1\ldots a_k},v^i_{b_1\ldots b_k})$,
$k=0,\ldots,m$. These transition functions
fulfill the cocycle relations. There is the exact sequence over $Z$ of vector
bundles
\be
0\to \w E^*\op\ot_Z\pr_2VE\to\cP_E\to \w E^*\op\ot_Z TZ\to 0. 
\ee
It is readily observed that every linear connection 
\beq
\G=dz^A\ot (\dr_A +\G_A{}^a{}_bv^b\dr_a) \label{cmp71}
\eeq
on the vector bundle $E\to Z$ yields the splitting 
\beq
\G_S:u^A\dr_A \mapsto u^A(\dr_A +\G_A{}^a{}_bc^b) \label{cmp70}
\eeq
of this exact sequence and the corresponding decomposition
\be  
u= u^A\dr_A + u^a\dr_a=u^A(\dr_A +\G_A{}^a{}_bc^b\dr_a) + (u^a-
u^A\G_A{}^a{}_bc^b)\dr_a
\ee
of sections of $\cP_E$. 
One can think of $\G_S$ as being a superconnection, but this is not a
connection on the fiber bundle $\cP_E\to Z$ in the conventional sense. The
sheaf of sections of
$\cP_E\to Z$ is isomorphic to the sheaf $\der\,\cB$. Global sections of the
vector bundle
$\cP_E\to Z$ constitute the $\cB^0$-module of supervector fields on $Z$ which
is also a Lie
superalgebra with respect to the bracket 
\be
[u,u']=uu' + (-1)^{\nw u\nw{u'}+1}u'u.
\ee

Similarly, the $\cB^0$-dual $\cP^*_E$ of $\cP_E$ is a vector bundle over $Z$
which is locally isomorphic to the vector bundle
\be
\cP^*_E\mid_U\approx \w E^*\op\ot_Z(\pr_2VE^*\op\oplus_Z T^*Z)\mid_U,
\ee
and has the transition functions
\be
&& v'_{j_1\ldots j_kj}= \rho^{-1}{}_{j_1}^{a_1}\cdots
\rho^{-1}{}_{j_k}^{a_k} \rho^{-1}{}_j^a v_{a_1\ldots a_ka}, \nonumber\\
&& z'_{i_1\ldots i_kA}=
\rho^{-1}{}_{i_1}^{b_1}\cdots
\rho^{-1}{}_{i_k}^{b_k}\left[z_{b_1\ldots b_kA}+ \frac{k!}{(k-1)!} 
v_{b_1\ldots b_kj}\dr_A(\rho^j_{b_k})\right] 
\ee
of the bundle coordinates $(z_{a_1\ldots a_kA},v_{b_1\ldots b_kj})$,
$k=0,\ldots,m$, with respect to the dual bases $\{dz^A\}$ in $T^*Z$ and
$\{dc^b\}$ in $\pr_2V^*E=E^*$. There is the exact sequence
\beq
0\to \w E^*\op\ot_ZT^*Z\to\cP^*_E\to \w E^*\op\ot_Z \pr_2VE^*\to 0.
\label{cmp72}
\eeq
The sheaf of sections of $\cP^*_E\to Z$ is isomorphic to
the sheaf $\der^*\cB$. Global sections of the vector bundle
$\cP^*\to Z$ constitute the $\cB^0$-module of exterior 1-superforms 
\beq
\f=\f_A dz^A + \f_adc^a\label{z790}
\eeq
on $Z$ with the coordinate transformation law
\be
\f'_a=\rho^{-1}{}_a^b\f_b, \qquad \f'_A=\f_A
+\rho^{-1}{}_a^b\dr_A(\rho^a_j)\f_bc^j.
\ee
The superform (\ref{z790}) is homogeneous iff
\be
\nw{\f_A}=\nw{\f_B}=\nw{\f_a}+1=\nw{\f_b}+1=\nw\f, 
\qquad \forall A,B,a,b. 
\ee
Then
the morphism (\ref{z789}) can be seen as the interior product 
\beq
u\rfloor \f=u^A\f_A + (-1)^{\nw{\f_a}}u^a\f_a. \label{cmp65}
\eeq

Similarly to (\ref{cmp70}), every linear connection $\G$
(\ref{cmp71}) on the vector bundle $E\to Z$ yields the splitting of the exact
sequence (\ref{cmp72}) and the corresponding decomposition 
\be
\f=\f_A dz^A + \f_adc^a =(\f_A+\G_A{}^a{}_b\f_ac^b)dz^A +\f_a(dc^a
-\G_A{}^a{}_bc^bdz^A) 
\ee
of 1-superforms on $Z$.

As in the non-graded case, by exterior $k$-superforms $\f$ are meant sections
of the graded exterior product $\ol\w^k_Z\cP^*_E$ 
which is the quotient of the tensor product
$\ot^k_Z\cP_E^*$
with respect to its subbundle generated by elements
\be
\varphi\ot \varphi'+(-1)^{[\varphi][\varphi']}\varphi'\ot\varphi, \qquad
\varphi,\varphi'\in\cP^*_E.
\ee
It is readily observed that $k$-superforms, $k=0,1,\ldots$, constitute a
$(\bZ,\bZ_2)$-bi-graded ring $\cB^*$ with respect to the graded
exterior product such that
\be
 \f\ol\w\si =(-1)^{\nm\f\nm\si +\nw\f\nw\si}\si\ol\w \f.  
\ee
The interior product (\ref{cmp65})
is extended to the ring $\cB^*$ by the rule  
\be
u\rfloor (\f\ol\w\si)=(u\rfloor \f)\ol\w \si
+(-1)^{\nm\f+\nw\f\nw{u}}\f\ol\w(u\rfloor\si). 
\ee

Using (\ref{cmp50}), one can introduce the graded exterior differential of
superfunctions in accordance with the condition 
\beq
u\rfloor df=u(f) \label{cmp67}
\eeq
for an arbitrary supervector field $u$ (this condition differs from that of
\cite{bart}). The graded differential is extended uniquely to the ring
$\cB^*$ of superforms by the familiar rules
\be
d(\f\ol\w\si)= (d\f)\ol\w\si +(-1)^{\nm\f}\f\ol\w(d\si), \qquad  dd=0.
\ee
It takes the coordinate form
\be
d\f= dz^A \ol\w \dr_A(\f) +dc^a\ol\w \dr_a(\f), 
\ee
where the left derivatives 
$\dr_A$, $\dr_a$ act on the coefficients of superforms by the rule
(\ref{cmp50}), and are graded commutative with the forms $dz^A$, $dc^a$.
With the interior product and the exterior graded differential, the Lie
derivative of a superform $\f$ along a supervector field $u$ can be given by
the familiar formula
\beq
\bL_u\f= u\rfloor d\f + d(u\rfloor\f). \label{cmp66}
\eeq
If $f$ is a superfunction, we obtain from (\ref{cmp67}) and (\ref{cmp66})
that $\bL_uf=u(f)$ as usual.

\end{document}